\documentclass[a4paper,12pt]{article}
\usepackage{amssymb,amsmath}
\usepackage{epsfig,fancyhdr}

\usepackage{a4}
\usepackage{parskip}
\usepackage{color,soul,rotating}
\usepackage{amsthm,amsxtra,amscd,relsize,multirow}
\usepackage[all,2cell]{xy}\UseAllTwocells

\newcommand{\sect}[1]{\setcounter{equation}{0}\section{#1}}

\def\a{\alpha}

\def\vf{\varphi}

\def\o{\omega}

\def\cosh{\mathrm{cosh}}

\def\p{\partial}
\def\rb{\right}
\def\lb{\left}
\def\axs{AdS_5\times S^5}
\newcommand{\eq}[1]{\begin{equation} #1 \end{equation}}
\newcommand{\al}[1]{\begin{align} #1 \end{align}}
\newcommand{\ml}[1]{\begin{multline} #1 \end{multline}}

\def\cp{CP^3}

\begin{document}
\begin{titlepage}
\markright{\bf TUW--10--20}
\title{On semiclassical calculation of three-point functions in $AdS_4\times\cp$}

\author{D.~Arnaudov${}^{\star}$ and R.~C.~Rashkov${}^{\dagger,\star}$\thanks{e-mail:
rash@hep.itp.tuwien.ac.at.}
\ \\ \ \\
${}^{\star}$  Department of Physics, Sofia
University,\\
5 J. Bourchier Blvd, 1164 Sofia, Bulgaria
\ \\ \ \\
${}^{\dagger}$ Institute for Theoretical Physics, \\ Vienna
University of Technology,\\
Wiedner Hauptstr. 8-10, 1040 Vienna, Austria
}
\date{}
\end{titlepage}

\maketitle
\thispagestyle{fancy}

\begin{abstract}
Recently there has been progress on the computation of two- and three-point correlation functions with
two ``heavy'' states via semiclassical methods. We extend this analysis to the case of $AdS_4\times\cp$,
and examine the suggested procedure for the case of several simple string solutions.
By making use of AdS/CFT duality, we derive the relevant correlation functions of operators belonging to
the dual ${\cal N}=6$ Chern-Simons gauge theory.
\end{abstract}

\sect{Introduction}

The attempts to establish a correspondence between the large $N$
limit of gauge theories and string theory has more than 30 years history and
 over the years it showed different faces. Recently an explicit realization of
this correspondence was provided by the Maldacena conjecture about AdS/CFT correspondence
\cite{Maldacena}. The convincing results from the duality between type IIB string
theory on $AdS_5\times S^5$ and ${\cal N}=4$ super Yang-Mills theory~\cite{Maldacena,GKP,Witten} made
this subject a major research area, and many
fascinating new features have been established.

An important part of the current understanding of the duality between gauge theories and strings (M-theory)
is the world-volume dynamics of the branes. Recently there has been a number of works focused on the understanding of
the world-volume dynamics of multiple M2-branes -- an interest inspired by the investigations of Bagger, Lambert
and Gustavsson \cite{Bagger:2006sk}, based on the structure of Lie 3-algebra.

Motivated by the possible description of the world-volume
dynamics of coincident membranes in M-theory, a new class of
conformal invariant, maximally supersymmetric field theories in 2+1
dimensions has been found \cite{Schwarz:2004yj,ABJM}.
The main feature of these theories is that they contain gauge fields with Chern-Simons-like  kinetic
terms. Aharony, Bergman, Jafferis and Maldacena \cite{ABJM} proposed a new gauge/string duality between an
${\cal N}=6$ superconformal Chern-Simons theory (ABJM theory) coupled with bi-fundamental matter, describing
$N$ membranes on $S^7/\mathbb Z_k$, and M-theory on $AdS_4\times S^7/\mathbb Z_k$.

The ABJM theory actually consists of two Chern-Simons theories of level $k$
and $-k$ correspondingly, and each with gauge group $SU(N)$. The two pairs of
chiral superfields transform in the bi-fundamental representations
of $SU(N)\times SU(N)$, and the R-symmetry is $SU(4)$ as it should be for
 $\mathcal N=6$ supersymmetry. It was observed in
\cite{ABJM} that there exists a natural definition of a 't Hooft coupling $\lambda = N/k$.
It was noticed that in the 't Hooft limit $N\rightarrow \infty$ with $\lambda$ held fixed,
 one has a continuous coupling  $\lambda$, and the ABJM theory is weakly coupled for $\lambda \ll 1$.
The ABJM theory is conjectured to be dual to M-theory on $AdS_4\times S^7 / \mathbb Z_k$ with
$N$ units of four-form flux. In the scaling limit $N, k \rightarrow\infty$
with $k \ll N \ll k^5$ the theory can be compactified to type IIA string theory on $AdS_4\times P^3$.
Thus, the AdS/CFT correspondence, which has led to many exciting developments in the duality between type IIB string
theory on $AdS_5\times S^5$ and ${\cal N}=4$ super Yang-Mills theory, is now being extended to AdS${}_4$/CFT${}_3$,
and is expected to constitute a new example of exact gauge/string theory duality.

The semiclassical strings have played, and still play, an important role in studying various aspects of AdS${}_5$/SYM${}_4$ correspondence \cite{Bena:2003wd}-\cite{Alday:2005gi}. The development in this
subject gives a strong hint about how the new emergent duality can be investigated. An important
role in these studies is played by the integrability. The superstrings on $AdS_4\times\cp$ as a coset were first
studied in \cite{Arutyunov:2008if}\footnote{See also \cite{Stefanski:2008}.}, which opened the door for
investigation of the integrable structures in the theory. To pursue these issues it is necessary to have the
complete superstring action. It was noticed that
the string supercoset model does not describe the entire dynamics of type IIA superstring in $AdS_4\times\cp$,
but only its
subsector. The complete string dual of the ABJM model, i.e. the
complete type-IIA Green-Schwarz string action in  $AdS_4\times\cp$ superspace,
has been constructed in \cite{Gomis:2008jt}.

Various properties on the gauge theory side and tests on the string theory side
as rigid rotating strings, pp-wave limit, relation to spin chains, and certain
limiting cases, as well as pure spinor formulation  have been considered
\cite{Arutyunov:2008if}-\cite{Zarembo:2009au}. In these intensive studies many impressive results were obtained, but still the understanding of
this duality is far from complete.

Recently, there has been progress in the semiclassical calculation of two- and three-point correlation
functions for the case of $\axs$ \cite{Janik:2010gc}-\cite{Georgiou:2010}. Motivated by these studies,
we consider the correlation functions of two ``heavy'' and one ``light'' operators in the ABJM theory.
To compute the correlation functions by making use of AdS/CFT we consider several simple solutions and apply
the methods for calculation of three-point correlators suggested in \cite{Janik:2010gc}-\cite{Georgiou:2010}.
Although the identification of the gauge theory operators corresponding to particular solutions is in general
a complicated task, we use the simplest known examples of such solutions and find the correlation functions.
Certainly this study should be considered as a first step towards applying ideas and techniques
known from the better investigated case of AdS${}_5$/CFT${}_4$ duality.

The paper is organized as follows. In the Introduction we briefly present the basic facts about ABJM theory.
To explain the method, in the next Section we give a short review of the procedure for calculating
semiclassically 3-point correlation functions for the case of generic backgrounds. Next, we proceed with
the calculation of correlation functions on $AdS_4\times\cp$. We conclude with a brief discussion on the results.

\paragraph{ABJM and strings on $AdS_4\times \cp$}

To find the ABJM theory one starts with analysis of the M2-brane dynamics governed by the 11-dimensional
supergravity action \cite{ABJM}
\eq{
S=\frac{1}{2\kappa_{11}^2}\int dx^{11}\sqrt{-g}\left(R-\frac{1}{2\cdot
4!}F_{\mu\nu\rho\sigma}F^{\mu\nu\rho\sigma}\right)-\frac{1}{12\kappa_{11}^2}
\int C^{(3)}\wedge F^{(4)}\wedge F^{(4)},
\label{abjm-1}
}
where $\kappa_{11}^2=2^7\pi^8l_p^9$ with $l_p$ being the 11-dimensional Planck length. Solving for the
equations of motion
\eq{
R^\mu_\nu=\frac{1}{2}\left(\frac{1}{3!}F^{\mu\a\beta\gamma}F_{\nu\a\beta\gamma}
-\frac{1}{3\cdot 4!}\delta^\mu_\nu
F_{\a\beta\rho\sigma}F^{\a\beta\rho\sigma}\right),
\label{abjm-2}
}
and
\eq{
\p_{\sigma}(\sqrt{-g}F^{\sigma\mu\nu\xi})=\frac{1}{2\cdot
(4!)^2}\varepsilon^{\mu\nu\xi\a_1\dots\a_8}F_{\a_1\dots\a_4}F_{\a_5\dots\a_8}\,,
\label{abjm-3}
}
one can find the M2-brane solutions whose near horizon limit becomes $AdS_4\times S^7$
\eq{
ds^2=\frac{R^2}{4}ds^2_{AdS_4}+R^2 ds^2_{S^7}.
\label{abjm-4}
}
In addition, we have $N'$ units of four-form flux
\eq{
 F^{(4)}=\frac{3R^3}{8}\varepsilon_{AdS_4}, \quad R=l_p(2^5 N'\pi^2)^{\frac{1}{6}}.
\label{abjm-5}
}
Next we consider the quotient $S^7/\mathbb Z_k$
acting as $z_i\rightarrow e^{i\frac{2\pi}{k}}z_i$. It is convenient
first to write the metric on $S^7$ as
\eq{
ds^2_{S^7}=(d\varphi'+\omega)^2 + ds^2_{\cp},
\label{abjm-6}
}
where
\al{
&ds^2_{\cp}=\frac{\sum_idz_id\bar z_i}{r^2} - \frac{|\sum_iz_id\bar z_i|^2}{r^4},
\quad r^2\equiv\sum_{i=1}^4|z_i|^2,\notag\\
&d\varphi'+\omega\equiv\frac{i}{2r^2}(z_id\bar z_i-\bar z_idz_i),\quad d\omega=J=id\left(\frac{z_i}{r}\right)
d\left(\frac{\bar z_i}{r}\right),
\label{abjm-7}
}
and then to perform the $\mathbb Z_k$ quotient identifying
$\vf'=\vf/k$ with $\vf\sim \vf+2\pi$ ($J$ is proportional to the K\"ahler form on $\cp$).
The resulting metric becomes
\eq{
ds^2_{S^7/{\mathbb Z}_k}=\frac{1}{k^2}(d\vf+k\o)^2 + ds^2_{\cp}.
}
One can observe that the first volume factor on the right hand side is divided by factor of $k$ compared to
the initial one. In order to have consistent quantized flux one must impose $N'=kN$, where $N$ is
the number of quanta of the flux on the quotient. We note that the spectrum of the supergravity fields of
the final theory is just the projection of the
initial $AdS_4\times S^7$ onto the $\mathbb Z_k$ invariant states. In this setup there is a natural
definition of {}'t~Hooft coupling $\lambda\equiv N/k$. Decoupling limit should be taken
as $N,k\rightarrow \infty$, while $N/k$ is kept fixed.

One can follow now \cite{ABJM} to make reduction to type IIA with the following final result
\al{
ds^2_{string}&=\frac{R^3}{k}\left(\frac{ds^2_{AdS_4}}{4}+ds^2_{\cp}\right),\\
e^{2\phi}&=\frac{R^3}{k^3}\sim\frac{N^{1/2}}{k^{5/2}}=\frac{1}{N^2}\left(\frac{N}{k}\right)^{5/2}\!\!\!\!\!\!,\\
F_{4}&=\frac38R^3\varepsilon_4,\quad F_2=kd\omega=kJ\,.
}
We end up then with $AdS_4\times\cp$ compactification of type IIA string theory with $N$ units
of $F_4$ flux on $AdS_4$ and $k$ units of $F_2$ flux on the $CP^1\subset\cp$ 2-cycle. Since we are interested
in semiclassical
fundamental string solutions in this background, the RR fluxes will play no role here. The radius of curvature
in string units is $R^2_{str}=\frac{R^3}{k}=2^{5/2}\pi\sqrt{\lambda}$.

On the boundary side we have a three-dimensional ${\cal N}=6$ superconformal theory \cite{ABJM}, which is a
Chern-Simons theory with gauge group
$SU(N)\times SU(N)$ with bifundamental superfields $A_1, A_2$ and
anti-bifundamental superfields $B_1, B_2$. The action has a pure
Chern-Simons part
\eq{
S_{CS}=\frac{k}{4\pi}\int
\left(A_{(1)}\wedge dA_{(1)}+\frac23 A_{(1)}^3-A_{(2)}\wedge dA_{(2)}- \frac23 A_{(2)}^3\right),
}
and superpotential
\eq{
W=\frac{4\pi}{k}{\rm Tr}\left(A_1B_1A_2B_2-A_1B_2A_2B_1\right).
}
Note that the signs before the Chern-Simons terms are opposite and
the superpotential has actually an $SU(2)\times SU(2)$ global
symmetry, which acts on the $A$'s and the $B$'s separately.
The levels of the Chern-Simons theory for the two components of the gauge group are $k$ and $-k$,
respectively. When $k$ and $N$ satisfy $k\ll N\ll k^5$, this field
theory is dual to the IIA superstring theory on $AdS_4\times\cp$ that we have described above.

\sect{Calculation of three-point functions}

Let us consider backgrounds of the form $AdS_{d+1}\times Y$, where $Y$ is a compact manifold. We will use Poincare
coordinates $(z,x)$ for the $AdS$ part, so the boundary will be a $d$-dimensional Minkowski space with
coordinates $x$. As was shown in \cite{Costa:2010} the three-point correlation function at strong coupling of
two ``heavy'' and one ``light'' boundary operators assumes the following form
\eq{
\left\langle{\cal O}_A(x_i){\cal O}_A(x_f){\cal D}_\chi(y)\right\rangle\approx\frac{I_\chi[\bar{X}(\tau,\sigma),
\bar{s};y]}{|x_i-x_f|^{2\Delta_A}}\,,
}
where $\Delta_A$ is the scaling dimension of the ``heavy'' operators, and with a slight abuse of notation
\eq{
I_\chi[\bar{X},\bar{s};y]=i\int_{-\bar{s}/2}^{\bar{s}/2}d\tau\int d\sigma\left.
\frac{\delta S_{P}[\bar{X},\bar{s},\Phi]}{\delta\chi}\right|_{\Phi=0}\,K_\chi(\bar{X};y)\,.
\label{Ichi}
}
Here $\Phi$ denotes the supergravity fields, one of which is $\chi$. It sources the operator ${\cal D}_\chi$
near the boundary. Also, the Polyakov action in conformal gauge is
\eq{
S_P[\bar{X},\bar{s},\Phi]=-g\int_{-\bar{s}/2}^{\bar{s}/2}d\tau\int d\sigma e^{\phi/2}\eta^{\alpha\beta}
\partial_\alpha\bar{X}^A\partial_\beta\bar{X}^Bg_{AB}+\cdots,
\label{Polyakov}
}
where $g$ is the coupling, $\phi$ is the dilaton field, and $g_{AB}$ is the background metric. In addition,
$\bar{s}$ is the saddle-point value of the modular parameter $s$ on the worldsheet cylinder, whose minimization
of area gives the string propagator (2-point function). $\bar{X}$ stands for a classical solution to the equations of
motion, which corresponds to an operator with large quantum numbers in the dual gauge theory, namely ${\cal O}_A$.
In what follows, we will confine ourselves to solutions which are point-particle in $AdS$. They are described with
\begin{align}
\label{xzparticle}
z&=z(\tau)=R/\cosh\kappa\tau\,,\nonumber\\
x&=x(\tau)=R\tanh\kappa\tau+x_0\,.
\end{align}
As was shown in \cite{Janik:2010gc}
\eq{
\label{kappabc}
\kappa\approx\frac{2}{s}\log{\frac{x_f}{\varepsilon}}\,,
}
where $\varepsilon$ is a regulator and the $\kappa$ is defined through $t=\kappa\tau$.
From now on, for simplicity, we assume $\chi$ to be the dilaton. Thus, ${\cal D}_\chi$ is equal
 to the Lagrangian density ${\cal L}$, and the bulk-to-boundary propagator in \eqref{Ichi} has the following
form \cite{Freedman:1998}
\eq{
K_\chi(\bar{X};y)=K_{\phi}(z(\tau),x(\tau);y)=\frac{\Gamma(d)}{\pi^{d/2}\Gamma(d/2)}\left(\frac{z(\tau)}{z^2(\tau)
-(x(\tau)-y)^2}\right)^d\!\!.
}

The detailed analysis in \cite{Janik:2010gc} shows that there is a subtlety in obtaining the string propagator,
so that the
classical solution for the cylinder coincides with the classical state. Therefore, when calculating $\bar{s}$
we have to consider (this does not apply to point-like strings)
\begin{equation}
\tilde{S}_P=S_P-\int_{-s/2}^{s/2}d\tau\int d\sigma\,\Pi^A\dot{X}_A\,.
\label{NewAction}
\end{equation}

Also, the following relation is expected to hold (analyzed for $d=4$ in \cite{Costa:2010,Roiban:2010})
\eq{
\left\langle{\cal O}_A(0){\cal O}_A(x_f){\cal L}(y)\right\rangle\approx-\frac{\Gamma(d/2)}{2\pi^{d/2}}
\frac{g^2\partial\Delta_A}{\partial g^2}\frac{x_f^{d-2\Delta_A}}{y^d(x_f-y)^d}\,.
\label{aLAA}
}
For more details we refer to \cite{Costa:2010,Zarembo:2010,Roiban:2010}.

\sect{Three-point functions in $AdS_4\times\cp$}

In this Section we apply the method described in the previous Section to the case of
strings in $AdS_4\times\cp$ background.

In order to fix the notation, we write down the explicit form of the metric \cite{Cvetic:2000}
\begin{multline}
ds^2_{AdS_4\times\cp}/R_{str}^2=\frac14 \frac{dz^2+dx^2}{z^2}+\left(d\xi^2+\cos^2\xi\sin^2\xi
(d\psi+\frac12\cos\theta_1d\varphi_1-\frac12\cos\theta_2d\varphi_2)^2\right.\\
\left.+\frac14\cos^2\xi(d\theta_1^2+\sin^2\theta_1
d\varphi_1^2)+\frac14\sin^2\xi(d\theta_2^2+\sin^2\theta_2 d\varphi_2^2)\right),
\label{metric}
\end{multline}
where the ranges of the coordinates are
$$0\leq z,\,\,x\in\mathbb{M}^{(1,2)},\,\,0\leq\xi\leq\pi/2,\,\,-2\pi\leq\psi\leq2\pi,\,\,0\leq\theta_i\leq\pi,
\,\,0\leq\varphi_i\leq2\pi,\,i=1,2\,.$$

If we assume the point-particle solution in $AdS_4$ \eqref{xzparticle}, the Polyakov action in conformal gauge
can be written as
\begin{multline}
S_P[X,s,\Phi]=\frac{g}{4}\int_{-s/2}^{s/2}d\tau\int d\sigma e^{\phi/2}\biggl[\kappa^2 +
4\dot{\xi}^2 - 4{\xi'}^2 + \sin^22\xi(\dot{\psi}^2-{\psi'}^2)\\
+ \cos^2\xi(\dot{\theta}_1^2-{\theta'_1}^2) + \sin^2\xi(\dot{\theta}_2^2-{\theta'_2}^2) + \sin^22\xi
\left(\cos\theta_1(\dot{\psi}\dot{\varphi_1}-\psi'\varphi'_1)-\cos\theta_2(\dot{\psi}\dot{\varphi_2}-\psi'\varphi'_2)
\right)\\
+ \cos^2\xi(\sin^2\theta_1+\cos^2\theta_1\sin^2\xi)(\dot{\varphi_1}^2-{\varphi'_1}^2) + \sin^2\xi(\sin^2\theta_2
+\cos^2\theta_2\cos^2\xi)(\dot{\varphi_2}^2-{\varphi'_2}^2)\\
- \frac12\sin^22\xi\cos\theta_1\cos\theta_2(\dot{\varphi_1}\dot{\varphi_2}-\varphi'_1\varphi'_2)\biggr],
\end{multline}
where $g=\frac{\sqrt{\tilde{\lambda}}}{4\pi},\,\tilde{\lambda}^2=32\pi^2\lambda^2$ in the case of $AdS_4\times\cp$.

The Killing vectors along the isometric coordinates give the conserved quantities
\begin{eqnarray}
E\!\!\!&=&\!\!\!\frac{g\kappa_{spin}}{2}\int d\sigma,\quad\kappa_{spin}=-i\kappa>0\,,\\
J_{\psi}\!\!\!&=&\!\!\!\frac{g}{2}\int d\sigma\sin^22\xi\left(\dot{\psi}+\frac{\cos\theta_1}{2}\dot{\varphi_1}
-\frac{\cos\theta_2}{2}\dot{\varphi_2}\right)\!,\\
J_1\!\!\!&=&\!\!\!\frac{g}{2}\int d\sigma\left[\cos^2\xi\sin^2\theta_1\dot{\varphi_1}+\sin^22\xi
\left(\frac{\cos^2\theta_1}{4}\dot{\varphi_1}
+\frac{\cos\theta_1}{2}(\dot{\psi}-\frac{\cos\theta_2}{2}\dot{\varphi_2})\right)\right]\!,\\
J_2\!\!\!&=&\!\!\!\frac{g}{2}\int d\sigma\left[\sin^2\xi\sin^2\theta_2\dot{\varphi_2}+\sin^22\xi
\left(\frac{\cos^2\theta_2}{4}\dot{\varphi_2}
-\frac{\cos\theta_2}{2}(\dot{\psi}+\frac{\cos\theta_1}{2}\dot{\varphi_1})\right)\right]\!.
\end{eqnarray}
The equations of motion for $\psi, \varphi_1, \varphi_2$ are just the conservation of $J_{\psi}, J_1,
 J_2$.

To find the general solution is a hard task, but one can derive specific solutions to the string equations of
motion. Then one can calculate the dispersion relations, which by making use of AdS/CFT correspondence give the
anomalous dimensions of certain gauge theory operators. To this date one can compute the three-point
correlation functions for two ``heavy'' and one ``light'' operators. Thus, our strategy for what follows will be to find
particular string solutions, identify the gauge theory operators, and use the solutions to calculate the
correlators.

\subsection{Point-like string}

Let us consider as a warm up example the limit of point-like string with mass $m$ \cite{Janik:2010gc,Costa:2010}. On
dimensional grounds it can be inferred that $m\sim g^{1/2}$. The particle action takes the following form
\eq{
S_P[X,s,\Phi=0]=\frac12\int_{-s/2}^{s/2}d\tau\left(\kappa^2 - m^2\right)=\frac12\left(\frac{4}{s^2}\log^2
\frac{x_f}{\varepsilon} - m^2\right)s\,.
}
The saddle point in the modular parameter is
\eq{
\label{saddleparticle}
\bar{s}=-\frac{2i}{m}\log{\frac{x_f}{\varepsilon}}\,.
}
Consequently,
\eq{
e^{iS_P[\bar{X},\bar{s},\Phi=0]}=\left(\frac{\varepsilon}{x_f}\right)^{2m}.
}
This gives the correct dimension in the large $\Delta_A$ limit, for which $\Delta_A\approx m$.

To compute the 3-point function, we need to evaluate
\al{
I_\phi[\bar{X},s;y]&=\frac{i}{2\pi^2}\int_{-s/2}^{s/2}d\tau\left(\kappa^2 - m^2\right)\left(
\frac{z}{z^2+(x-y)^2}\right)^3\nonumber\\
&=\frac{is}{64\pi\log\frac{x_f}{\varepsilon}}\left(\frac{4}{s^2}\log^2\frac{x_f}{\varepsilon} - m^2\right)
\frac{x_f^3}{y^3(x_f-y)^3}\,.
}
At the modular parameter saddle point \eqref{saddleparticle}, this expression becomes simply
\eq{
I_\phi[\bar{X},\bar{s};y]=-\frac{m}{16\pi}\frac{x_f^3}{y^3(x_f-y)^3}\,.
}
We conclude that in the large $\Delta_A$ limit
\eq{
\langle{\cal O}_A(0){\cal O}_A(x_f){\cal L}(y)\rangle\approx-\frac{\Delta_A}{16\pi}
\frac{x_f^{3-2\Delta_A}}{y^3(x_f-y)^3}\,.
}
It can be seen that indeed \eqref{aLAA} holds, provided that $\Delta_A\approx m\sim g^{1/2}$.

\subsection{Circular string I}

We start the analysis of less trivial examples with the case of circular strings.
To find a particular solution we make the ansatz \cite{Chen:2008qq}
\eq{
\xi=\xi_0={\rm const},\,\theta_1=0,\,\theta_2=n\sigma,\,\psi=\omega_1\tau,\,\varphi_1
=\omega_2\tau,\,\varphi_2=0\,,
}
which can be easily checked to be a solution to the equations of motion.
Then, simple arguments give
\eq{
\cos2\xi_0=-\frac{n^2}{(2\omega_1+\omega_2)^2}\,.
}
It is straightforward to find the conserved quantities and the dispersion relation for this
string solution:
\eq{
J_{\psi}=\pi g(\omega_1+\frac12\omega_2)\sin^22\xi_0=2J_1,\,J_2=0\,,
}
and
\eq{
E^2=\frac{J_{\psi}^2}{\sin^22\xi_0}+\pi^2g^2n^2\sin^2\xi_0\,.
}

Let us apply now the procedure outlined briefly in the previous Section. For the Polyakov
action in this case we have
\begin{align}
S_P[\bar{X},s,\Phi=0]&=\frac{g}{4}\int_{-s/2}^{s/2}d\tau\int_{0}^{2\pi}d\sigma\left(\kappa^2
+ \sin^22\xi_0(\dot{\psi}+\frac{1}{2}\dot{\varphi_1})^2 - \sin^2\xi_0\theta^{\prime2}_2\right)\nonumber\\
&=\frac{\pi g}{2}\left(\frac{4}{s^2}\log^2\frac{x_f}{\varepsilon} + (\omega_1+\frac{1}{2}\omega_2)^2\sin^22\xi_0
- n^2\sin^2\xi_0\right)s\,,
\end{align}
where the relation \eqref{kappabc} is used.
The modified action \eqref{NewAction} in this case takes the form
\eq{
\tilde{S}_P[\bar{X},s,\Phi=0]=\frac{\pi g}{2}\left(\frac{4}{s^2}\log^2\frac{x_f}{\varepsilon} -
(\omega_1+\frac{1}{2}\omega_2)^2\sin^22\xi_0 - n^2\sin^2\xi_0\right)s\,.
}
The saddle point with respect to the modular parameter $s$ is given by
\eq{
\label{saddlecircularI}
\bar{s} =-\frac{2i}{\sqrt{(\omega_1+\frac{1}{2}\omega_2)^2\sin^22\xi_0+n^2\sin^2\xi_0}}
\log{\frac{x_f}{\varepsilon}}\,.
}
On the other hand, looking at \eqref{kappabc}, this implies the Virasoro constraint
$$
\kappa=i\sqrt{(\omega_1+\frac{1}{2}\omega_2)^2\sin^22\xi_0 + n^2\sin^2\xi_0}\,.
$$
Therefore, the evaluation of $e^{i\tilde{S}_P[\bar{X},\bar{s},\Phi=0]}$ at the saddle point gives
\eq{
e^{i\tilde{S}_P[\bar{X},\bar{s},\Phi=0]}=\left(\frac{\varepsilon}{x_f}\right)^{2\pi g\sqrt{(\omega_1
+\frac{1}{2}\omega_2)^2\sin^22\xi_0 + n^2\sin^2\xi_0}}.
}
This expression tells us that the dimension $\Delta_A = \sqrt{\frac{J_{\psi}^2}{\sin^22\xi_0} + \pi^2g^2n^2\sin^2\xi_0} = E$ in agreement
with the AdS/CFT prediction.

Now we turn to the derivation of the 3-point function. First we evaluate
\begin{align}
I_\phi[\bar{X},s;y]&=\frac{ig}{2\pi^2}\int_{-s/2}^{s/2}d\tau\!\int_{0}^{2\pi}\!d\sigma\left(\kappa^2
+ \sin^22\xi_0(\dot{\psi}+\frac{1}{2}\dot{\varphi_1})^2 - \sin^2\xi_0\theta^{\prime2}_2\right)\!
\left(\frac{z}{z^2+(x-y)^2}\right)^3\nonumber\\
&=\frac{igs}{32\log\frac{\varepsilon}{x_f}}\left(\frac{4}{s^2}\log^2\frac{x_f}{\varepsilon} +
(\omega_1+\frac{1}{2}\omega_2)^2\sin^22\xi_0 - n^2\sin^2\xi_0\right)\frac{x_f^3}{y^3(x_f-y)^3}\,.
\end{align}
At the saddle point \eqref{saddlecircularI}, we get
\eq{
I_\phi[\bar{X},\bar{s};y] =-\frac{\pi g^2n^2\sin^2\xi_0}{8E}\frac{x_f^3}{y^3(x_f-y)^3}\,.
}
Therefore, the final expression for the three-point correlation function is
\eq{
\langle{\cal O}_A(0){\cal O}_A(x_f){\cal L}(y)\rangle\approx-\frac{\pi g^2n^2\sin^2\xi_0}{8E}
\frac{x_f^{3-2E}}{y^3(x_f-y)^3}\,.
}
It can be seen that indeed \eqref{aLAA} holds, provided that $J_{\psi}$ is kept constant.

\subsection{Circular string II}

The next example of a circular string solution is \cite{Chen:2008qq}
\eq{
\xi=\xi_0={\rm const},\,\theta_1=n_1\sigma,\,\theta_2=n_2\sigma,\,\psi=\omega_1\tau,\,\varphi_1=0,\,
\varphi_2=0\,.
}
Then
\eq{
\cos2\xi_0=\frac{n_1^2-n_2^2}{4\omega_1^2}\,.
}
The charges and the dispersion relation for this solution are found to be
\eq{
J_{\psi}=\pi g\omega_1\sin^22\xi_0, J_1=J_2=0\,,
}
and
\eq{
E^2=\frac{J_{\psi}^2}{\sin^22\xi_0}+\pi^2g^2(n_1^2\cos^2\xi_0+n_2^2\sin^2\xi_0)\,.
}

Let us apply now the method used above to obtain the correlation function. First of all, we have
to evaluate the Polyakov action on this solution
\begin{align}
S_P[\bar{X},s,\Phi=0]&=\frac{g}{4}\int_{-s/2}^{s/2}d\tau\int_{0}^{2\pi}d\sigma\left(\kappa^2 +
\sin^22\xi_0\dot{\psi}^2 - \cos^2\xi_0\theta^{\prime2}_1 - \sin^2\xi_0\theta^{\prime2}_2\right)\nonumber\\
&=\frac{\pi g}{2}\left(\frac{4}{s^2}\log^2\frac{x_f}{\varepsilon} + \omega_1^2\sin^22\xi_0 - n_1^2\cos^2\xi_0
- n_2^2\sin^2\xi_0\right)s\,.
\end{align}
Note that we have used above the relation \eqref{kappabc}.
The modified action \eqref{NewAction} in this case becomes
\eq{
\tilde{S}_P[\bar{X},s,\Phi=0]=\frac{\pi g}{2}\left(\frac{4}{s^2}\log^2\frac{x_f}{\varepsilon} -
\omega_1^2\sin^22\xi_0 - n_1^2\cos^2\xi_0 - n_2^2\sin^2\xi_0\right)s\,.
}
The saddle point in the modular parameter $s$ is found to be
\eq{
\label{saddlecircularII}
\bar{s}=-\frac{2i}{\sqrt{\omega_1^2\sin^22\xi_0+n_1^2\cos^2\xi_0+n_2^2\sin^2\xi_0}}
\log{\frac{x_f}{\varepsilon}}\,.
}
Using \eqref{kappabc}, this implies the Virasoro constraint
$$
\kappa=i\sqrt{\omega_1^2\sin^22\xi_0+n_1^2\cos^2\xi_0+n_2^2\sin^2\xi_0}\,.
$$
It is easy now to calculate the contribution at the saddle point
\eq{
e^{i\tilde{S}_P[\bar{X},\bar{s},\Phi=0]}=\left(\frac{\varepsilon}{x_f}\right)^{2\pi g
\sqrt{\omega_1^2\sin^22\xi_0+n_1^2\cos^2\xi_0+n_2^2\sin^2\xi_0}},
}
which again gives the correct dimension $\Delta_A = \sqrt{\frac{J_{\psi}^2}{\sin^22\xi_0}+\pi^2g^2(n_1^2\cos^2\xi_0
+n_2^2\sin^2\xi_0)} = E$.

To obtain the three-point function we first evaluate
\begin{align}
&I_\phi[\bar{X},s;y]=\nonumber\\
&=\frac{ig}{2\pi^2}\int_{-s/2}^{s/2}d\tau\int_{0}^{2\pi}d\sigma\left(\kappa^2 + \sin^22\xi_0\dot{\psi}^2 -
 \cos^2\xi_0\theta^{\prime2}_1 - \sin^2\xi_0\theta^{\prime2}_2\right)\left(\frac{z}{z^2+(x-y)^2}\right)^3\nonumber\\
&=\frac{igs}{32\log\frac{\varepsilon}{x_f}}\left(\frac{4}{s^2}\log^2\frac{x_f}{\varepsilon} +
\omega_1^2\sin^22\xi_0 - n_1^2\cos^2\xi_0 - n_2^2\sin^2\xi_0 \right)\frac{x_f^3}{y^3(x_f-y)^3}\,.
\end{align}
At the saddle point \eqref{saddlecircularII} we have
\eq{
I_\phi[\bar{X},\bar{s};y]=-\frac{\pi g^2(n_1^2\cos^2\xi_0+n_2^2\sin^2\xi_0)}{8E}\frac{x_f^3}{y^3(x_f-y)^3}\,.
}
Therefore, for the three-point function we find
\eq{
\langle{\cal O}_A(0){\cal O}_A(x_f){\cal L}(y)\rangle\approx-\frac{\pi g^2(n_1^2\cos^2\xi_0
+n_2^2\sin^2\xi_0)}{8E}\frac{x_f^{3-2E}}{y^3(x_f-y)^3}\,.
}
It can be seen that indeed \eqref{aLAA} holds, provided that $J_{\psi}$ is kept constant.

For the particular case of $n_1=\pm n_2, \xi_0=\pi/4$ we obtain
\eq{
\langle{\cal O}_A(0){\cal O}_A(x_f){\cal L}(y)\rangle\approx-\frac{E^2-J_{\psi}^2}{8\pi E}
\frac{x_f^{3-2E}}{y^3(x_f-y)^3}\,.
}
In this case, the field theory dual is the operator ${\rm Tr}\left((A_1B_1)^J(A_2B_2)^J\right)$ \cite{Chen:2008qq}.
It has $J_{\psi}=2J,\,J_1=J_2=0$, and at the classical level $E=2J$.

\subsection{Folded string I}

Another class of relevant string solutions is the case of folded onto themselves strings.
To find such string configurations we make the ansatz \cite{Chen:2008qq}
\eq{
\xi=\xi(\sigma),\,\theta_1=\theta_2=0,\,\psi=\omega_1\tau,\,\varphi_1=\omega_2\tau,\,\varphi_2=\omega_3\tau.
}
One can easily show that this ansatz is consistent with the equations of motion.
The only nontrivial equation to be solved is
\eq{
\xi^{\prime\prime}=-\frac{\tilde{\omega}^2}{4}\sin4\xi\,,
}
where $\tilde{\omega}=\omega_1+(\omega_2-\omega_3)/2$.
The conserved charges for this ansatz are
\eq{
J_{\psi}=\frac{g\tilde{\omega}}{2}\int_0^{2\pi}d\sigma\sin^22\xi,\quad J_1=J_{\psi}/2,\quad J_2
=-J_1,
}
and the conserved quantities are expressed through elliptic integrals in the following way
\eq{
E=2g\sin2\xi_0K(q), J_{\psi}=2g(K(q)-E(q))\,.
\label{foldedIcons}
}
Here $q=\sin^22\xi_0$, and  $\xi_0$ is the maximal value of $\xi$.
The equations \eqref{foldedIcons} give the dispersion
relation in parametric form.

Let us turn now to the correlation function. First we have to evaluate the Polyakov action
on this solution
\begin{align}
S_P[\bar{X},s,\Phi=0]&=\frac{g}{4}\int_{-s/2}^{s/2}d\tau\int_{0}^{2\pi}d\sigma\left(\kappa^2 - 4\xi^{\prime2}
+ \tilde{\omega}^2\sin^22\xi\right)\nonumber\\
&=\frac{g}{4}\int_{0}^{2\pi}d\sigma\left(\frac{4}{s^2}\log^2\frac{x_f}{\varepsilon} - 4\xi^{\prime2} +
\tilde{\omega}^2\sin^22\xi\right)s\,.
\end{align}
Note that above we have used the relation \eqref{kappabc}.
The modified action \eqref{NewAction} in this case becomes
\eq{
\tilde{S}_P[\bar{X},s,\Phi=0]=\frac{g}{4}\int_{0}^{2\pi}d\sigma\left(\frac{4}{s^2}\log^2\frac{x_f}{\varepsilon}
 - 4\xi^{\prime2} - \tilde{\omega}^2\sin^22\xi\right)s\,.
}
The saddle point with respect to the modular parameter $s$ is given by
\eq{
\label{saddlefoldedI}
\bar{s}=-\frac{2i}{\sqrt{4\xi^{\prime2}+\tilde{\omega}^2\sin^22\xi}}\log{\frac{x_f}{\varepsilon}}\,,
}
which by means of \eqref{kappabc} implies the Virasoro constraint $\kappa=i\sqrt{4\xi^{\prime2}+
\tilde{\omega}^2\sin^22\xi}$. The saddle point contribution is then
\eq{
e^{i\tilde{S}_P[\bar{X},\bar{s},\Phi=0]}=\left(\frac{\varepsilon}{x_f}\right)^{g\int_{0}^{2\pi}
d\sigma\sqrt{4\xi^{\prime2}+\tilde{\omega}^2\sin^22\xi}},
}
which formally gives again the correct dimension $\Delta_A = \frac{g}{2}\int_{0}^{2\pi}d\sigma\sqrt{4\xi^{\prime2}
+\tilde{\omega}^2\sin^22\xi} = E$.

Analogously to the previous cases we obtain
\eq{
I_\phi[\bar{X},\bar{s};y]=-\frac{1}{8\pi}\left(E-\frac{J_{\psi}}{\sin2\xi_0}\right)\frac{x_f^3}{y^3(x_f-y)^3}\,.
}
Therefore, the final expression we find is
\eq{
\langle{\cal O}_A(0){\cal O}_A(x_f){\cal L}(y)\rangle\approx-\frac{1}{8\pi}
\left(E-\frac{J_{\psi}}{\sin2\xi_0}\right)\frac{x_f^{3-2E}}{y^3(x_f-y)^3}\,.
}
It can be seen with a little effort that \eqref{aLAA} holds, provided that $J_{\psi}$ is kept constant. If we take the large $J$ limit, which corresponds to $\xi_0\rightarrow\pi/4$ \cite{Chen:2008qq}, we have $E\rightarrow\infty$, $J_{\psi}\rightarrow\infty$ and $E-J_{\psi}\sim 2g$. Also
\eq{
\langle{\cal O}_A(0){\cal O}_A(x_f){\cal L}(y)\rangle\approx-\frac{E-J_{\psi}}{8\pi}
\frac{x_f^{3-2E}}{y^3(x_f-y)^3}\,.
}

\subsection{Folded string II}

Let us consider another folded string extended along $\theta_1$ direction. The ansatz for this string configuration is \cite{Chen:2008qq}
\eq{
\xi=\pi/4,\,\theta_1=\theta_1(\sigma)=\pm\theta_2,\,\psi=\omega_1\tau,\,\varphi_1=\omega_2\tau,\,\varphi_2
=-\omega_2\tau.
\label{folded2-ans}
}
Simple check shows that \eqref{folded2-ans} solves the equations of motion provided
\eq{
\theta_1^{\prime\prime}=\omega_1\omega_2\sin\theta_1
}
is satisfied.

For this ansatz, there are two subcases which we will consider separately.

\paragraph{The case of $\omega_1=0$}\ \\

For this solution the charges and the dispersion relation are
\eq{
J_{\psi}=0,\,J_1=-J_2=\frac{\pi g\omega_2}{2}\,,
}
and
\eq{
E^2=4J_1^2+\pi^2g^2n^2.
}
Although this folded string solution is quite similar to some of the others, its field theory duals are actually very different \cite{Chen:2008qq}. Matching the quantum numbers, one obtains for gauge theory duals the following operators ${\rm Tr}\left((A_1B_1)^{J_1/2}(B_2^\dagger A_2^\dagger)^{J_1/2}\right)$. The energy
of the string in the large $J$ limit is
\eq{
E=2J_1+\frac{\pi^2g^2n^2}{4J_1}+\cdots.
}
Classically, the dual operator has dimension $\Delta_A=2J_1$, in consistency with the zero order string
energy. It should be kept in mind that the first order correction to the string
energy is of order $g^2/J_1$, similarly to the circular string II we discussed above.

Now we start with the derivation of the correlation function. To this end, we have to evaluate
the Polyakov action on the choice made above
\eq{
S_P[\bar{X},s,\Phi=0]=\frac{g}{4}\int_{-s/2}^{s/2}d\tau\int_{0}^{2\pi}d\sigma\left(\kappa^2 - \theta^{\prime2}_1
+ \omega_2^2\right)=\frac{\pi g}{2}\left(\frac{4}{s^2}\log^2\frac{x_f}{\varepsilon} - n^2 + \omega_2^2\right)s\,,
}
where we have imposed the relation \eqref{kappabc}.
The modified action \eqref{NewAction} needed for the correlation function in this case is
\eq{
\tilde{S}_P[\bar{X},s,\Phi=0]=\frac{\pi g}{2}\left(\frac{4}{s^2}\log^2\frac{x_f}{\varepsilon} - n^2 -
\omega_2^2\right)s\,.
}
Next, the saddle point with respect to the modular parameter $s$ is given by
\eq{
\label{saddlefoldedIIa}
\bar{s}=-\frac{2i}{\sqrt{n^2+\omega_2^2}}\log{\frac{x_f}{\varepsilon}}\,.
}
Having in mind \eqref{kappabc}, this implies the Virasoro constraint $\kappa=i\sqrt{n^2+\omega_2^2}$.
Putting things together, at the saddle point we find
\eq{
e^{i\tilde{S}_P[\bar{X},\bar{s},\Phi=0]}=\left(\frac{\varepsilon}{x_f}\right)^{2\pi g\sqrt{n^2+\omega_2^2}}.
}
This is in perfect agreement with AdS/CFT correspondence, i.e.
 the correct dimension is $\Delta_A = \sqrt{4J_1^2+\pi^2g^2n^2} = E$.

Now we turn to the 3-point function. First we evaluate
\begin{align}
I_\phi[\bar{X},s;y]&=\frac{ig}{2\pi^2}\int_{-s/2}^{s/2}d\tau\int_{0}^{2\pi}d\sigma\left(\kappa^2 - n^2 +
\omega_2^2\right)\left(\frac{z}{z^2+(x-y)^2}\right)^3\nonumber\\
&=\frac{igs}{32\log\frac{\varepsilon}{x_f}}\left(\frac{4}{s^2}\log^2\frac{x_f}{\varepsilon} - n^2 +
 \omega_2^2\right)\frac{x_f^3}{y^3(x_f-y)^3}\,,
\end{align}
then we substitute for the modular parameter its saddle point value \eqref{saddlefoldedIIa}. Thus
\eq{
I_\phi[\bar{X},\bar{s};y]=-\frac{E^2-4J_1^2}{8\pi E}\frac{x_f^3}{y^3(x_f-y)^3}\,.
}
The final expression we find is
\eq{
\langle{\cal O}_A(0){\cal O}_A(x_f){\cal L}(y)\rangle\approx-\frac{E^2-4J_1^2}{8\pi E}
\frac{x_f^{3-2E}}{y^3(x_f-y)^3}\,.
}
It can be seen that indeed \eqref{aLAA} gives correct answer, provided $J_1$ is kept constant.

\paragraph{The case of $\omega_1\neq0$}\ \\

For this solution the charges are a little bit more involving
\eq{
J_{\psi}=\frac{g}{2}\int_0^{2\pi}d\sigma(\omega_1+\omega_2\cos\theta_1),\,J_1=
\frac{g}{4}\int_0^{2\pi}d\sigma(\omega_2+\omega_1\cos\theta_1),\,
J_2=-J_1.
}
The conserved quantities can be easily expressed in terms of elliptic integrals in the following way
\eq{
J_{\psi}=\frac{2g}{\sqrt{-\omega_1\omega_2}}\left((\omega_1-\omega_2)K(x)+2\omega_2E(x)\right)\!, J_1=
\frac{g}{\sqrt{-\omega_1\omega_2}}\left((\omega_2-\omega_1)K(x)+2\omega_1E(x)\right),
}
where $x=\sin^2\frac{\theta_1(0)}{2}$. Using the Virasoro constraints and the periodicity condition one
obtains the relation
\eq{
\sqrt{-\o_1\o_2}=\frac{2}{\pi}K(x).
}
The dispersion relation is expressed in an implicit form as
\al{
&\lb(\frac{2E}{K(x)}\rb)^2-\lb(\frac{J_1+2J_2}{E(x)}\rb)^2=64g^2x\,,\label{dispf2-a}\\
&\lb(\frac{J_1-2J_2}{E(x)-K(x)}\rb)^2-\lb(\frac{J_1+2J_2}{E(x)}\rb)^2=64g^2.\label{dispf2-b}
}
The Polyakov action evaluated on this ansatz is
\begin{align}
S_P[\bar{X},s,\Phi=0]&=\frac{g}{4}\int_{-s/2}^{s/2}d\tau\int_{0}^{2\pi}d\sigma\left(\kappa^2 + \omega^2_1 +
\omega^2_2 + 2\omega_1\omega_2\cos\theta_1 - \theta_1^{'2}\right)\nonumber\\
&=\frac{g}{4}\int_{0}^{2\pi}d\sigma\left(\frac{4}{s^2}\log^2\frac{x_f}{\varepsilon} + \omega^2_1 + \omega^2_2 +
2\omega_1\omega_2\cos\theta_1 - \theta_1^{'2}\right)s\,,
\end{align}
where the relation \eqref{kappabc} was taken into account.
The modified action \eqref{NewAction} in this case takes the form
\eq{
\tilde{S}_P[\bar{X},s,\Phi=0]=\frac{g}{4}\int_{0}^{2\pi}d\sigma\left(\frac{4}{s^2}\log^2\frac{x_f}{\varepsilon} -
 \omega^2_1 - \omega^2_2 - 2\omega_1\omega_2\cos\theta_1 - \theta_1^{'2}\right)s\,.
}
To obtain the saddle point with respect to the modular parameter $s$ it is not necessary to evaluate the integral. Thus, one obtains
\eq{
\label{saddlefoldedIIb}
\bar{s}=-\frac{2i}{\sqrt{\omega^2_1+\omega^2_2+2\omega_1\omega_2\cos\theta_1+\theta_1^{'2}}}
\log{\frac{x_f}{\varepsilon}}\,.
}
As usual, \eqref{kappabc} implies the Virasoro constraint $\kappa=i\sqrt{\omega^2_1+\omega^2_2+
2\omega_1\omega_2\cos\theta_1+\theta_1^{'2}}\,.$ Then, at the saddle point we have the contribution
\eq{
e^{i\tilde{S}_P[\bar{X},\bar{s},\Phi=0]}=\left(\frac{\varepsilon}{x_f}\right)^
{g\int_{0}^{2\pi}d\sigma\sqrt{\omega^2_1+\omega^2_2+2\omega_1\omega_2\cos\theta_1+\theta_1^{'2}}},
}
which gives the correct dimension $\Delta_A = \frac{g}{2}\int_{0}^{2\pi}d\sigma\sqrt{\omega^2_1+\omega^2_2+
2\omega_1\omega_2\cos\theta_1+\theta_1^{'2}} = E$.

To derive the three-point function, first we evaluate
\begin{align}
I_\phi[\bar{X},s;y]&=\frac{ig}{2\pi^2}\int_{-s/2}^{s/2}d\tau\int_{0}^{2\pi}d\sigma\left(\kappa^2 + \omega^2_1 +
\omega^2_2 + 2\omega_1\omega_2\cos\theta_1 - \theta_1^{'2}\right)\left(\frac{z}{z^2+(x-y)^2}\right)^3\nonumber\\
&=\frac{igs}{64\pi\log\frac{\varepsilon}{x_f}}\int_{0}^{2\pi}d\sigma\left(\frac{4}{s^2}\log^2\frac{x_f}{\varepsilon} +
\omega^2_1 + \omega^2_2 + 2\omega_1\omega_2\cos\theta_1 - \theta_1^{'2}\right)\frac{x_f^3}{y^3(x_f-y)^3}\,.
\end{align}
The next step is to evaluate the above expression at the saddle point \eqref{saddlefoldedIIb}, i.e.
\eq{
I_\phi[\bar{X},\bar{s};y]=-\frac{E^2-\pi g(\omega_1J_{\psi}+2\omega_2J_1)}{8\pi E}
\frac{x_f^3}{y^3(x_f-y)^3}\,.
}
With all these at hand we find for the correlation function the expression
\eq{
\langle{\cal O}_A(0){\cal O}_A(x_f){\cal L}(y)\rangle\approx-\frac{E^2-\pi g(\omega_1J_{\psi}+
2\omega_2J_1)}{8\pi E}\frac{x_f^{3-2E}}{y^3(x_f-y)^3}\,.
}
It can be seen with a little more effort that indeed \eqref{aLAA} gives correct answer, provided $J_{\psi}$ and $J_1$ are kept constant.

To match the above angular momenta, the following identification was proposed in \cite{Chen:2008qq}
\eq{
\left\{
\begin{array}{ll}
{\rm Tr}\left((A_1B_1)^{\frac{J_{\psi}}{2}+J_1}(A_2B_2)^{\frac{J_{\psi}}{2}-J_1}
\right),&\hspace{3ex}\mbox{for $\omega_1+\omega_2>0$}\,,\\
{\rm Tr}\left((B_1^\dagger A_1^\dagger)^{-(\frac{J_{\psi}}{2}+J_1)}(A_2B_2)^{\frac{J_{\psi}}{2}-
J_1}\right),&\hspace{3ex}\mbox{for $\omega_1+\omega_2<0$}\,.
\end{array}
\right.
}
However, if one takes $\omega_1=-\omega_2$, the above operators reduce to ${\rm Tr}\left((A_2B_2)^{J_{\psi}}\right)$,
which is a BPS primary and has $\Delta=J_{\psi}$ without quantum correction, which is different from what we know from
the string calculation. Since the dispersion relations \eqref{dispf2-a} and \eqref{dispf2-b} are the same as
in the $AdS_5\times S^5$ case, it is natural to accept that point of view, namely, that there exists an interpolating
function governing the anomalous dimension transition from $g^2$ in the weak coupling regime
to $g$ in the strong coupling one. Certainly this point needs more work and clarifications.

\subsection{Giant magnon and spiky string I}

Two of the very important for the AdS/CFT correspondence classes of string solutions are the so called
giant magnon and spiky string solutions. The former have the shape of arcs, while the latter
are wound strings developing a single spike.
In this subsection we will consider some of the simplest such solutions
and will calculate their three-point correlation functions.
We start with the ansatz suggested in \cite{Ryang:2008rc}
\eq{
\xi=\xi(\eta),\,\theta_1=\theta_2=\frac{\pi}{2},\,\psi=0,\,\varphi_1=\omega_1\tau+f_1(\eta),\,\varphi_2=
\omega_2\tau+f_2(\eta)\,,
\label{mag-spi-ans}
}
where $\eta=\alpha\sigma+\beta\tau$.
For this ansatz, the conserved quantum numbers we have are
\ml{
\Delta\varphi_2=-\int d\varphi_2=\frac{\alpha\beta}{\alpha^2-\beta^2}\int_{-\infty}^{\infty}d\sigma
\left(\frac{\kappa_{spin}^2}{\omega_2\sin^2\xi}-\omega_2\right),\,
J_{\psi}=0\,,\\
J_1=\frac{g\alpha^2\omega_1}{2(\alpha^2-\beta^2)}\int_{-\infty}^{\infty}d\sigma\cos^2\xi,\,
J_2=\frac{g}{2(\alpha^2-\beta^2)}\int_{-\infty}^{\infty}d\sigma\sin^2\xi\left(\alpha^2\omega_2-
\frac{\beta^2\kappa_{spin}^2}{\omega_2\sin^2\xi}\right),
}
where $\Delta\varphi_2$ is the angle deficit due to the specific string shape.
The dispersion relation can be obtained to be
\eq{
E-J_2=\sqrt{J_1^2+4g^2\sin^2\frac{p}{4}}
}
for magnons ($p=\Delta\varphi_2$), and
\eq{
E-\frac{g\Delta\varphi_2}{2}=2g\arcsin\frac{\sqrt{J_2^2-J_1^2}}{2g}
}
for spiky strings.

Substituting the ansatz \eqref{mag-spi-ans} into the Polyakov action we find
\begin{align}
&S_P[\bar{X},s,\Phi=0]=\nonumber\\
&=\frac{g}{4}\int_{-s/2}^{s/2}d\tau\int_{-\infty}^{\infty}d\sigma\left(\kappa^2 + 4(\dot{\xi}^2-\xi^{\prime2}) +
\cos^2\xi(\dot{\varphi_1}^2-\varphi_1^{\prime2}) + \sin^2\xi(\dot{\varphi_2}^2-\varphi_2^{\prime2})\right)\nonumber\\
&=\frac{g}{4}\int_{-\infty}^{\infty}d\sigma\left(\frac{4}{s^2}\log^2\frac{x_f}{\varepsilon} +
4(\dot{\xi}^2-\xi^{\prime2}) + \cos^2\xi(\dot{\varphi_1}^2-\varphi_1^{\prime2}) + \sin^2\xi(\dot{\varphi_2}^2-
\varphi_2^{\prime2})\right)s\,,
\end{align}
where as before, we used the relation \eqref{kappabc}.
The modified action \eqref{NewAction} in this case we find to be
\begin{align}
&\tilde{S}_P[\bar{X},s,\Phi=0]=\nonumber\\
&=\frac{g}{4}\int_{-\infty}^{\infty}d\sigma\left(\frac{4}{s^2}\log^2\frac{x_f}{\varepsilon} - 4(\dot{\xi}^2+
\xi^{\prime2}) - \cos^2\xi(\dot{\varphi_1}^2+\varphi_1^{\prime2}) - \sin^2\xi(\dot{\varphi_2}^2+
\varphi_2^{\prime2})\right)s\nonumber\\
&=\frac{g}{4\alpha}\int_{-\infty}^{\infty}d\eta\left(\frac{4}{s^2}\log^2\frac{x_f}{\varepsilon} -
4(\alpha^2+\beta^2)\xi^{\prime2} - \cos^2\xi[(\omega_1+\beta f_1^{\prime})^2+\alpha^2f_1^{\prime2}]\right.
\nonumber\\
&- \left.\sin^2\xi[(\omega_2+\beta f_2^{\prime})^2+\alpha^2f_2^{\prime2}]\right)s\,,
\end{align}
where prime denotes derivative with respect to $\eta$ in the last two lines. The saddle point
with respect to the modular parameter $s$ is given by
\eq{
\label{saddlemagnonI}
\bar{s}=-\frac{2i}{\sqrt{4(\alpha^2+\beta^2)\xi^{\prime2}+\cos^2\xi[(\omega_1+\beta f_1^{\prime})^2+
\alpha^2f_1^{\prime2}]+\sin^2\xi[(\omega_2+\beta f_2^{\prime})^2+\alpha^2f_2^{\prime2}]}}
\log{\frac{x_f}{\varepsilon}}\,,
}
which again due to \eqref{kappabc} implies the Virasoro constraint
$$
\kappa=i\sqrt{4(\alpha^2+\beta^2)\xi^{\prime2}+\cos^2\xi[(\omega_1+\beta f_1^{\prime})^2+
\alpha^2f_1^{\prime2}]+\sin^2\xi[(\omega_2+\beta f_2^{\prime})^2+\alpha^2f_2^{\prime2}]}\,.
$$
Combining all these we find that at the saddle point the contribution of
$e^{i\tilde{S}_P[\bar{X},\bar{s},\Phi=0]}$ is
\eq{
e^{i\tilde{S}_P[\bar{X},\bar{s},\Phi=0]}=\left(\frac{\varepsilon}{x_f}\right)^{2E}.
}
We note that this gives the correct relation between the dimension of the gauge theory operator and the
energy, $\Delta_A = E$.

In order to obtain the 3-point function we evaluate
\begin{align}
I_\phi[\bar{X},s;y]&=\frac{ig}{2\pi^2}\int_{-s/2}^{s/2}d\tau\int_{-\infty}^{\infty}d\sigma\left(\kappa^2 +
4(\dot{\xi}^2-\xi^{\prime2})\right.\nonumber\\
&+ \left.\cos^2\xi(\dot{\varphi_1}^2-\varphi_1^{\prime2}) + \sin^2\xi(\dot{\varphi_2}^2-\varphi_2^{\prime2})
\right)\left(\frac{z}{z^2+(x-y)^2}\right)^3.
\end{align}
At the saddle point \eqref{saddlemagnonI} the above integral takes the form
\eq{
I_\phi[\bar{X},\bar{s};y]=-\frac{\alpha g}{16\pi\kappa_{spin}}\int_{-\infty}^{\infty}d\eta\left(4\xi^{\prime2} +
\cos^2\xi f_1^{\prime2} + \sin^2\xi f_2^{\prime2}\right)\frac{x_f^3}{y^3(x_f-y)^3}\,.
}
For the three-point function for both giant magnon and spiky strings we end up with the expression
\eq{
\langle{\cal O}_A(0){\cal O}_A(x_f){\cal L}(y)\rangle\approx-\frac{\kappa_{spin}E-\omega_1J_1-
\omega_2J_2}{8\pi\kappa_{spin}}\frac{x_f^{3-2E}}{y^3(x_f-y)^3}\,.
}
In the derivation we did not specify what type of string solutions we are using.
 Now we turn to the specific cases of magnons and spiky strings.

\paragraph{Giant magnon I}\ \\

The giant magnon solutions are singled out by the relation $\kappa_{spin}=\omega_2$. Then
\eq{
\langle{\cal O}_A(0){\cal O}_A(x_f){\cal L}(y)\rangle\approx-
\frac{(E-J_2)^2-J_1^2}{8\pi(E-J_2)}\frac{x_f^{3-2E}}{y^3(x_f-y)^3}\,.
}
It can be seen that indeed \eqref{aLAA} holds, provided that $p, J_1, J_2$ are kept constant.

\paragraph{Spiky string I}\ \\

The spiky strings are defined by $\kappa_{spin}=\frac{\alpha\omega_2}{\beta}$. Then
\eq{
\langle{\cal O}_A(0){\cal O}_A(x_f){\cal L}(y)\rangle\approx-\frac{\alpha EJ_2+
\beta(J_1^2-J_2^2)}{8\pi\alpha J_2}\frac{x_f^{3-2E}}{y^3(x_f-y)^3}\,.
}
It can be seen that indeed \eqref{aLAA} holds, provided that $\Delta\varphi_2, J_1,
J_2$ are kept constant.

\subsection{Giant magnon and spiky string II}

In this subsection we will consider more giant magnon and spiky string solutions for
strings extended in $\xi$ and $\psi$ directions.
The ansatz we are using to obtain such solutions is \cite{Ryang:2008rc}
\eq{
\xi=\xi(\eta),\,\theta_1=\theta_2=\frac{\pi}{2},\,\psi=\omega\tau+f(\eta),\,\varphi_1=\varphi_2=\varphi
=\nu\tau,
}
where $\eta=\alpha\sigma+\beta\tau$.
For this ansatz we have for the conserved quantities the expressions
\ml{
\Delta\psi=-\int d\psi=\frac{\alpha\beta}{\alpha^2-\beta^2}\int_{-\infty}^{\infty}d\sigma
\left(\frac{\kappa_{spin}^2-\nu^2}{\omega\sin^2\xi}-\omega\right),\\
J_{\psi}=\frac{g}{2(\alpha^2-\beta^2)}\int_{-\infty}^{\infty}d\sigma\sin^22\xi\left(\alpha^2\omega-
\frac{\beta^2(\kappa_{spin}^2-\nu^2)}{\omega\sin^22\xi}\right),\,
J_{\varphi}=\frac{\nu E}{\kappa_{spin}}\,.
}
The giant magnons possess the dispersion relation
\eq{
\sqrt{E^2-J_{\varphi}^2}-J_{\psi}=g\sin\frac{p}{2}\,,
}
where $p=\Delta\psi$. Analogously, the dispersion relation for spiky strings is
\eq{
\sqrt{E^2-J_{\varphi}^2}-\frac{g\Delta\psi}{2}=g\arcsin\frac{J_{\psi}}{g}\,.
}

We start the considerations with the Polyakov action, which for this ansatz takes the form
\begin{align}
&S_P[\bar{X},s,\Phi=0]=\nonumber\\
&=\frac{g}{4}\int_{-s/2}^{s/2}d\tau\int_{-\infty}^{\infty}d\sigma\left(\kappa^2 + 4(\dot{\xi}^2-\xi^{\prime2}) +
 \sin^22\xi(\dot{\psi}^2-\psi^{\prime2}) + (\dot{\varphi}^2-\varphi^{\prime2})\right)\nonumber\\
&=\frac{g}{4}\int_{-\infty}^{\infty}d\sigma\left(\frac{4}{s^2}\log^2\frac{x_f}{\varepsilon} +
4(\dot{\xi}^2-\xi^{\prime2}) + \sin^22\xi(\dot{\psi}^2-\psi^{\prime2}) + (\dot{\varphi}^2-\varphi^{\prime2})
\right)s\,.
\end{align}
The corresponding modified action \eqref{NewAction} we find to be
\begin{align}
&\tilde{S}_P[\bar{X},s,\Phi=0]=\nonumber\\
&=\frac{g}{4}\int_{-\infty}^{\infty}d\sigma\left(\frac{4}{s^2}\log^2\frac{x_f}{\varepsilon} -
 4(\dot{\xi}^2+\xi^{\prime2}) - \sin^22\xi(\dot{\psi}^2+\psi^{\prime2}) - (\dot{\varphi}^2+
\varphi^{\prime2})\right)s\nonumber\\
&=\frac{g}{4\alpha}\int_{-\infty}^{\infty}d\eta\left(\frac{4}{s^2}\log^2\frac{x_f}{\varepsilon} -
4(\alpha^2+\beta^2)\xi^{\prime2} - \sin^22\xi[(\omega+\beta f^{\prime})^2+\alpha^2f^{\prime2}] - \nu^2\right)s\,,
\end{align}
where prime denotes derivative with respect to $\eta$ in the last line. The saddle point in the modular
parameter $s$ is given by
\eq{
\label{saddlemagnonII}
\bar{s}=-\frac{2i}{\sqrt{4(\alpha^2+\beta^2)\xi^{\prime2}+\sin^22\xi[(\omega+\beta f^{\prime})^2+
\alpha^2f^{\prime2}]+\nu^2}}\log{\frac{x_f}{\varepsilon}}\,.
}
Again, having in mind \eqref{kappabc} the saddle point implies the Virasoro constraint
$$
\kappa=i\sqrt{4(\alpha^2+\beta^2)\xi^{\prime2}+\sin^22\xi[(\omega+\beta f^{\prime})^2+\alpha^2f^{\prime2}]+
\nu^2}\,.
$$
The contribution from the saddle point, which we have to substitute in the general expression, is
\eq{
e^{i\tilde{S}_P[\bar{X},\bar{s},\Phi=0]}=\left(\frac{\varepsilon}{x_f}\right)^{2E}.
}
This gives the correct dimension $\Delta_A = E$.

The next step is to evaluate the three-point correlation function.
In order to do that, first we evaluate
\begin{align}
I_\phi[\bar{X},s;y]&=\frac{ig}{2\pi^2}\int_{-s/2}^{s/2}d\tau\int_{-\infty}^{\infty}d\sigma\left(\kappa^2 +
4(\dot{\xi}^2-\xi^{\prime2})\right.\nonumber\\
&+ \left.\sin^22\xi(\dot{\psi}^2-\psi^{\prime2}) + \dot{\varphi}^2 - \varphi^{\prime2}\right)
\left(\frac{z}{z^2+(x-y)^2}\right)^3,
\end{align}
which at the saddle point \eqref{saddlemagnonII} gives
\eq{
I_\phi[\bar{X},\bar{s};y]=-\frac{\alpha g}{16\pi\kappa_{spin}}\int_{-\infty}^{\infty}d\eta\left(4\xi^{\prime2} +
\sin^22\xi f^{\prime2}\right)\frac{x_f^3}{y^3(x_f-y)^3}\,.
}
Analogously to the previous cases, we obtain for the three-point function
\eq{
\langle{\cal O}_A(0){\cal O}_A(x_f){\cal L}(y)\rangle\approx-\frac{E^2-J_{\varphi}^2-
\frac{\omega EJ_{\psi}}{\kappa_{spin}}}{8\pi E}\frac{x_f^{3-2E}}{y^3(x_f-y)^3}\,.
\label{gm+ssII}
}
Next, having the general expression \eqref{gm+ssII}, we have to distinguish between the giant magnon and spiky strings.

\paragraph{Giant magnon II}\ \\

The giant magnons are singled out by the relation $\kappa_{spin}^2-\nu^2=\omega_2$. Then
\eq{
\langle{\cal O}_A(0){\cal O}_A(x_f){\cal L}(y)\rangle\approx-\frac{g(J_{\psi}\sin\frac{p}{2}+
g\sin^2\frac{p}{2})}{8\pi E}\frac{x_f^{3-2E}}{y^3(x_f-y)^3}\,.
}
It can be seen that indeed \eqref{aLAA} holds, provided that $p, J_{\psi}, J_{\varphi}$ are kept constant.

\paragraph{Spiky string II}\ \\

The spiky strings are specified by  $\kappa_{spin}^2-\nu^2=\frac{\alpha^2\omega_2}{\beta^2}$. Then
\eq{
\langle{\cal O}_A(0){\cal O}_A(x_f){\cal L}(y)\rangle\approx-\frac{\alpha(E^2-J_{\varphi}^2)-
\beta J_{\psi}\sqrt{E^2-J_{\varphi}^2}}{8\pi\alpha E}\frac{x_f^{3-2E}}{y^3(x_f-y)^3}\,.
}
It can be seen that indeed \eqref{aLAA} holds, provided that $\Delta\psi, J_{\psi}, J_{\varphi}$ are kept constant.

To conclude this Section a few comments are in order. First of all, careful and at some points lengthy but
straightforward calculation (which we will skip here) shows that $-g^2\p\Delta_A/\p g^2$ reproduces
correctly the structure constants of the correlation functions we found. Secondly, as is well known, due to
large winding the dispersion relations for the spiky strings are quite different from those of the giant magnons. Nevertheless,
the anomalous dimensions as well as the couplings of the correlation functions are correctly reproduced.

\sect{Conclusion}

The AdS/CFT correspondence passed through many controversial developments over the last decade. The holographic
conjecture has been tested in many cases and impressive results about anomalous dimensions of the gauge theory
operators, integrable structures, etc., and crucial properties of various gauge theories at strong coupling have been
established. The main challenge ahead is to find efficient methods for calculation of the correlation functions.

Recently, a method for obtaining three-point correlation functions beyond the supergravity approximation
has been suggested \cite{Zarembo:2010,Costa:2010}\footnote{Such considerations were initiated in \cite{Janik:2010gc}. An approach
based on insertion of vertex operators was put forward in \cite{Roiban:2010}.}.
The authors consider string theory on $AdS_5\times S^5$ and calculate three-point correlation functions of
two heavy (string) and one light (supergravity) states at strong coupling.

In this paper we apply the ideas for calculation of correlation functions developed for AdS${}_5$/CFT${}_4$ to the
case of AdS${}_4$/CFT${}_3$ correspondence. As in the former case, the latter correspondence is supposed to be exact, and
it is reasonable to ask whether one can apply the same technique to obtain the three-point correlation functions.
We examine the method in the cases of various string solutions for which the spectroscopy of the anomalous dimensions
is elaborated. The string theory side computation of the correlation functions
reproduces the correct anomalous dimensions as predicted by AdS/CFT correspondence and the correct conformal
dependence on the three points at the boundary.

In most of the cases we found a good agreement with the expectation from AdS/CFT correspondence. The
dependence of the correlation functions on the boundary points is as predicted, i.e. the behavior is given by
$\Delta = E$, where $\Delta$ is the dimension of the gauge theory operator, and $E$ is the energy of
the corresponding string configuration.

There is another, equivalent in a way, approach to the problem we consider here, namely
using the vertex operators for the corresponding states \cite{Roiban:2010}. The careful analysis
states that the structure constants of the correlation functions are given by $-g^2\p\Delta_A/\p g^2$. We checked this relation for all the correlators and found
perfect agreement.

These considerations can be considered as a first step towards the extension of the analysis
developed in $AdS_5\times S^5$ to the case of another example of string/gauge theory duality,
namely $AdS_4\times CP_3$. Unfortunately, to the best of our knowledge, there are no known results
from gauge theory side for which our findings to be compared to. It would be interesting, however, to use
what is known from the Bethe ansatz to extract gauge theory results, which could confirm at least
some of our results. In any case, the correlation functions we obtained here should be considered only
as a first step. It would be interesting to extend these results to other light operators. It would also be interesting to see how some of the rigid properties of Chern-Simons theory can be translated to the string side.
Another direction would be to see what all these considerations mean from M-theory point of view, and especially the approach with vertex operator insertions.

\section*{Acknowledgments}
We would like to thank H. Dimov for valuable discussions and careful reading of the paper.
This work was supported in part by the Austrian Research Funds FWF P22000 and I192,
NSFB VU-F-201/06 and DO 02-257.


\end{document}